**Topological spin-torque diode effect in skyrmion-based magnetic tunnel junctions**

*Shuhui Liu*[†]*, Riccardo Tomasello*[†]*, Yuxuan Wu*[†]*, Bin Fang*[*]*, Aitian Chen, Dongxing Zheng, Baoshun Zhang, Emily Darwin, Hans J. Hug, Mario Carpentieri, Wanjun Jiang, Xixiang Zhang, Giovanni Finocchio*[*] *and Zhongming Zeng*[*]

S. Liu, Y. Wu, Dr. B. Fang[*], Prof. B. Zhang, Prof. Z. Zeng[*]

Nanofabrication facility, Suzhou Institute of Nano-Tech and Nano-Bionics, Chinese Academy of Sciences, Suzhou, Jiangsu 215123, China

School of Nano Technology and Nano Bionics, University of Science and Technology of China, Hefei, Anhui 230026, People's Republic of China

E-mail: bfang2013@sinano.ac.cn; zmzeng2012@sinano.ac.cn

Prof. G. Finocchio[*]

Department of Mathematical and Computer Sciences, Physical Sciences and Earth Sciences, University of Messina, I-98166, Messina, Italy

E-mail: gfinocchio@unime.it

Prof. R. Tomasello, Prof. M. Carpentieri

Department of Electrical and Information Engineering, Politecnico di Bari, I-70125 Bari, Italy

Prof. A. Chen, Dr. D. Zheng, Prof. X. Zhang

Physical Science and Engineering Division, King Abdullah University of Science and Technology, Thuwal 23955–6900, Saudi Arabia

Dr. E. Darwin, Prof. Hans J. Hug

Empa, Swiss Federal Laboratories for Materials Science and Technology, Ueberlandstrasse 129, 8600 Dübendorf, Switzerland

Prof. Hans J. Hug

Department of Physics, University of Basel, Klingelbergstrasse 82, 4056 Basel, Switzerland

Prof. W. Jiang
State Key Laboratory of Low-Dimensional Quantum Physics and Department of Physics, Tsinghua University, Beijing 100084, China.

[†]These authors contributed equally: Shuhui Liu, Riccardo Tomasello, Yuxuan Wu

**Abstract**

The growing market and massive use of Internet of Things nodes is placing unprecedented demands of energy efficient hardware for edge computing and microwave devices. In particular, magnetic tunnel junctions (MTJs), as main building blocks of spintronic microwave technology, can offer a path for the development of compact and high-performance microwave detectors. On the other hand, the fascinating field of skyrmionics is bridging together concepts from topology and spintronics. Here, we show the proof-of-concept of the direct electrical excitation and detection of the dynamics of a topological protected magnetic texture, i.e. skyrmion at room temperature and for a wide region of applied fields, including the zero-field case. This topological spin-torque diode is realized with an MTJ on top of a skyrmionic material. Quantitative Magnetic Force Microscopy measurements are employed to confirm the existence of a single skyrmion in the MTJ free layer. Spin-torque diode electrical measurements show the electrical excitation via spin-transfer torque (STT) of a skyrmion resonant mode with frequencies near 4 GHz and a selectivity one order of magnitude smaller than the uniform modes excited in the same device. Micromagnetic simulations identify these dynamics with the excitation of the breathing mode and point out the role of thickness dependent magnetic parameters (magnetic anisotropy field and Dzyaloshinkii–Moriya interaction) in both stabilizing and exciting the magnetic skyrmions. This work marks a milestone for the development of topological spin-torque diodes.

## 1. Introduction

The electrical manipulation and detection of topological quasi-particles and phases are key aspects that can add functionalities at device level, impacting the enhancement of several emerging fields in condensed matter physics[1-7].

In the last years, the interest on magnetic skyrmions and skyrmionic phases has grown thanks to the study of their fundamental properties and their potential as building blocks for applications in magnetic storage[8-13] and reservoir, probabilistic and neuromorphic computing[14-18]. The rapid advancements and optimization of material development and nanofabrication for controlling and manipulating magnetic skyrmions[19-21] have opened up an exciting research direction in spintronics, in which the skyrmion electrical detection can be achieved in magnetic tunnel junctions (MTJs) via the tunneling magnetoresistive signal (TMR) [22]. In the industry-compatible CoFeB/MgO/CoFeB MTJ stack, the TMR signal related to skyrmions stabilized in the CoFeB free layer (FL) was first indirectly measured at low temperature[23], and then at room temperature[24]. Furthermore, the efforts to develop MTJ stacks to simultaneously perform imaging and electrical detection of skyrmions was successfully proved in a CoFeB/MgO/Ta/Co MTJ with engineered Dzyaloshinkii-Moriya interaction (DMI), where a TMR signal of 1.5% was achieved[25]. Currently, the primary strategy employed to enhance the electrical signal of a skyrmion is to combine multilayers hosting room temperature magnetic skyrmions[26-29] with an optimized MTJ stack. This approach has been effective in demonstrating the electrical detection of skyrmions with a TMR larger than 20%[30-31], and the 100% TMR signature of a single skyrmion moving underneath an MTJ under the action of the current induced spin orbit torque[32]. The previous experimental efforts focused mainly on the static properties of skyrmions. However, one of the next breakthroughs, namely the development of skyrmion-based computing and microwave technology, calls for the electrical detection of the skyrmion dynamics in MTJs which relies on the simultaneous excitation and detection of a skyrmion mode. In this way, by combining industry-compatible MTJs and topologically non-trivial magnetization textures, we could introduce topological magnetism into realistic applications, i.e. realize topological devices. Perpendicular MTJs stand as the most promising venue to accomplish this task. On one hand, the spin-torque diode (STD) effect in MTJs has proven to be effective for computing applications[33-35]. On the other hand, theoretical predictions on topological STDs promise high performance in microwave detection and energy harvesting[36]. The latter exploits internal dynamics of the skyrmion known as breathing modes[37]. Differently from other internal modes, i.e. clockwise (CW) and counterclockwise (CCW), the breathing mode can be excited via an out-of-plane field[38] and it has only been experimentally observed with skyrmion lattices in bulk materials at low temperature[39-41]. For skyrmion lattices in magnetic multilayers at room temperature, only the CW and CCW modes have been excited by an in-plane ac field[42-43].

Here, we establish a multilayer combining a skyrmionic material hosting skyrmions at room temperature with a perpendicular MTJ stack on top, similarly to a previous work[30]. The MTJ is designed so that the magnetostatic field from the Skyrmionic Layer (SkyL) can transfer the skyrmion state into its FL. Quantitative Magnetic Force Microscopy (MFM) measurements confirm the existence of a single skyrmion in the MTJ FL. We have performed measurements of the spin-torque ferromagnetic resonance (ST-FMR) response as a function of the out-of-plane (OOP) field in MTJs with a diameter smaller than 300 nm. We have also achieved the zero-field excitation after having stabilized zero field skyrmion by means of the application of an ad hoc field protocol[44]. This is a scenario more relevant for applications. With full 3D micromagnetic simulations, we have identified the excitation of the skyrmion breathing mode as expected for MTJs with a perpendicular polarizer, which has a resonance frequency near 4 GHz. At large fields, we observe two modes, which are related to the excitation of uniform modes. Therefore, our work (i) demonstrates the first simultaneous electrical excitation via spin-transfer torque (STT) and detection of the experimental microwave response of a magnetic skyrmion, and (ii) represents the experimental proof-of-concept of a topological STD. These results pave the way for future developments of topological devices, not only for skyrmion-based microwave technology, but also for other applications, such as unconventional computing[45].

## 2. Results and Discussion

### 2.1. Sample description and characterization

The device stack is shown in **Figure 1**a. We use magnetron sputtering to grow magnetic films consisting of Ta (5)/CuN (20)/Ta (5)/[Pt (2.5)/Co (1)/Ta (0.5)]$_9$/Pt (2.5)/Co (1)/$Co_{40}Fe_{40}B_{20}$ (0.9)/MgO (0.85)/$Co_{20}Fe_{60}B_{20}$ (1.1)/Ta (0.5)/Co (0.3)/[Pt (1.5)/Co (0.4)]$_2$/Ru (0.85)/[Co (0.5)/Pt (1.5)]$_3$/Ru (5) – thicknesses in nm - on top of a thermally oxidized Si substrate. The $Co_{40}Fe_{40}B_{20}$ (0.9)/MgO (0.85)/$Co_{20}Fe_{60}B_{20}$ (1.1) is the part of the stack related to the MTJ (see Methods for more details on sample preparation). The top CoFeB is the perpendicularly magnetized reference layer (RL) pinned by a synthetic antiferromagnet (SAF) layer. The [Co (0.5)/Pt (1.5)]$_3$ acts as the hard layer (HL) and is used to enhance the coercive field. The bottom CoFeB acts as a FL and it is directly coupled with Pt (2.5)/Co (1)/Ta (0.5)]$_9$/Pt (2.5)/Co (1), i.e. the SkyL, which allows for stabilizing magnetic skyrmions at room temperature for a wide range of magnetic fields. Using electron-beam lithography and ion milling, these films were patterned into MTJ devices with circular cross sections of diameters ranging from 100 nm to 500 nm. **Figure 1**b shows the scanning electron microscope (SEM) image for an MTJ with a diameter of 270 nm. We have also characterized the stack cross section using transmitting electron microscopy (TEM) to evaluate the quality of magnetic films, as shown in **Figure 1**c, which exhibits the uniform layered structure and sharp interface characteristics (see Supplementary Figure S1 for additional images). The inset of **Figure 1**c provides a detailed view of the MgO layer and its two adjacent CoFeB layers, highlighting the high quality of the MTJ tunnel barrier giving rise to the TMR effect. Indeed, the MgO layer appears with a neat crystal phase arrangement.

**Figure 1**d shows an example of the OOP magnetic hysteresis loop for the thin films demonstrating the OOP easy axis, while the inset displays the minor loop with the change of the OOP field. The magnetic signal of the point A near zero field is marked by arrows of different colors, which correspond to that of the CoFeB/Ta/[Pt/Co/Ta]$_9$ multilayer. The magnetizations of the FL and the SkyL undergo antiferromagnetic coupling, as do the RL and the HL. The process of changing the relative magnetization direction between the FL and the RL is characterized by a step-shaped intermediate resistance state near zero magnetic field. To identify the magnetic configurations in this field region, we have performed magnetic imaging using polar magneto-optical Kerr effect (p-MOKE) microscopy (see the schematic of the measurement and the Kerr intensity magnetic-field loop in Supplementary Figure S2) in samples without the capping layer which are composed of the following stack [Ta(5)/MgO(0.85)/CoFeB(1.1)/Ta(0.5)/[Pt(2.5)/Co(1)/Ta(0.5)]$_9$. This is the same as the SkyL of the MTJ depicted in **Figure 1**a. When the external magnetic field is zero, the surface of the sample presents typical maze-shaped domains. With the gradual increasing of the magnetic field, the spin textures evolve from the maze-shaped domains into a configuration characterized by isolated skyrmions which micromagnetic simulations identify as pure Néel skyrmions. When the magnetic field is sufficiently large to saturate the magnetization, the skyrmions are annihilated (see Supplementary Figure S3 for a more detailed imaging of the field evolution of magnetic textures measured by p-MOKE). **Figure 1**e shows the magnetic state at 200 Oe where isolated skyrmions of around 250 nm diameter are observed. (the OOP magnetization components are color coded - white/black refers to +z/-z). We refer to this intermediate state, which hosts isolated magnetic skyrmions at room temperature for a wide range of magnetic fields, as a skyrmionic state. To verify experimentally whether the skyrmionic state in the SkyL is transferred into the CoFeB FL such that a skyrmion existing in the SkyL also exists in the adjacent CoFeB FL, we designed an experiment using two multilayers consisting of the MTJ stack up to the MgO layer but without the RL. In one multilayer sample, the CoFeB FL was present (**Figure 2**a), whereas it was replaced by a non-magnetic Tantalum (Ta) layer of the same thickness for the second multilayer sample (**Figure 2**b). Quantitative Magnetic force microscopy (MFM), performed under vacuum conditions with a high-quality factor cantilever to obtain an excellent signal-to-noise ratio even with a low moment super-sharp tip[46], was used to measure both samples under identical measurement conditions[47]. The applied OOP field was the maximum field possible before the skyrmion annihilated, $H = 1240$ Oe in **Figure 2**c and $H = 1180$ Oe in **Figure 2**d. Since the skyrmion cannot penetrate into the non-magnetic Ta layer, a contrast difference was expected between the skyrmions in the two samples only if the skyrmion was also stabilized into the CoFeB FL. **Figure 2**c shows the MFM frequency shift data of a skyrmion in the sample with the CoFeB FL, while **Figure 2**d shows the corresponding data of a skyrmion in the sample without the CoFeB FL. Both skyrmions have an apparent

diameter of about 65 nm. The frequency shift contrast of the sample with the CoFeB FL is about 6 Hz and considerably larger than the 4.8 Hz observed for the sample with the Ta layer, evident when comparing the data in **Figure 2**c and d and the skyrmion profiles in **Figure 2**e and f. The same contrast difference was also observed when measuring the samples in a domain state (not shown). This MFM results unequivocally confirm the presence of the skyrmion in both the SkyL and the CoFeB FL.

### 2.2. Topological spin-torque diode measurements

We next study the rectification characteristics of the MTJ device for the skyrmionic and uniform states. We focus on MTJs with a diameter of 270 nm which is comparable to the maximum size of the skyrmions observed by p-MOKE microscopy. We use the circuit shown in **Figure 3**a for the ST-FMR measurements (see Methods for more details). The signal generator and source meter generate a microwave current $I_{ac} = I_{ac,0}\sin(2\pi f_{ac}t + \varphi)$ of frequency $f_{ac}$, phase $\varphi$ and amplitude $I_{ac,0}$ and a direct current $I_{dc}$, respectively, which are applied to the MTJ device through a bias tee. We take advantage of the STD effect measuring the rectified voltage $V_{dc}$ across the MTJ as a function of the microwave frequency $f_{ac}$ to characterize the skyrmion modes. The rectified voltage depends on the amplitude of the microwave input, the amplitude of the MTJ oscillating resistance and the phase between them[48-49]. We have performed a systematic study of those modes as a function of the field amplitude and direction with respect to the out-of-plane axis characterized by the angle $\theta$ as defined in **Figure 3**a.

To identify the field region where the skyrmionic state is stable, we first measured the magnetoresistance as a function of the field amplitude for different field angles $\theta$. **Figure 3**b compares the results for $\theta$ = 0 and 30° (similar curves are observed for other field angles as illustrated in the Supplementary Figure S4a and S4b for $\theta$ = 20° and 40°). The direction of the RL is set along the -z direction, hence for negative fields the antiparallel state (AP) with a large resistance is stabilized. As the field increases, we observe a jump from the parallel (P) state to the skyrmionic state, similar to the process identified in **Figure 1**e. The offset in the hysteresis loop is given by the dipolar field from the SAF layer, which is not totally compensated. At large enough magnetic fields $H$ = 550 Oe, the AP state is stabilized. The reduction of the resistance of the AP state as the field keeps increasing is due to the rotation of the SAF layer magnetization, which, as can be observed, is a reversible process. A similar response is observed for the descending branch of the hysteresis loop while the magnetic field is decreasing. However, after the P state is stabilized the resistance remains constant up to the minimum field applied of -1000 Oe.

Then, we experimentally characterized the rectification voltage $V_{dc}$ by measuring the ST-FMR spectra as a function of the field amplitude for the ascending branch of the hysteresis loop. **Figure 3**c summarizes the data for fields larger than 250 Oe, $\theta = 30°$ and $P_{rf} = 5\mu W$. We identify two different regions, A and B. In region A, where the skyrmionic state is stable, a single mode with a frequency near 4 GHz is excited. Its frequency slightly decreases as a function of the field amplitude. **Figure 3**d shows an example of the rectified voltage as a function of the microwave frequency for $H$ = 260 Oe. It is characterized by a negative voltage peak and a narrow bandwidth. The fact that the MTJ FL is in the skyrmionic state and the magnetization of the RL is perpendicular allows us to assume that, in region A, the skyrmion breathing mode[36-37] is excited. As a first supporting experiment, we measured the ST-FMR in a conventional MTJ without the SkyL where the region A is not observed (see Supplementary Figure S5). Furthermore, micromagnetic simulations show the excitation of the breathing mode, as discussed later.

When the field increases, a transition from region A to B occurs. Here, two modes, a low frequency mode (LFM) and a high frequency mode (HFM) starting near 5 GHz and 8 GHz at $H$ = 600 Oe are detected. The frequency of the LFM (HFM) decreases (increases) with the field. **Figure 3**e displays the rectified voltage as a function of the microwave input frequency for $H$ = 950 Oe, as an example. In this field range, the MTJ is in the AP uniform state (see hysteresis loop in **Figure 3**b), therefore the HFM is related to the FL resonance mode, which is parallel to the applied field, while the LFM is linked to the RL resonance mode which is antiparallel to the field. Similar rectification curves are observed for different samples with the same nominal geometry and for a wide range input rf power up to 50 μW (see Supplementary Figure S6).

A direct comparison of the rectification curves for different modes shows that the maximum rectified voltage $V_{BM}$, indicated in **Figure 3**c and d, of the skyrmion breathing mode is smaller but of the same order of the one related to the excitation of the uniform mode. The resonant curve has a bandwidth of 163 MHz which is smaller as compared to 527 and 248 MHz for the HRM and LRM, respectively. This gives rise to a better selectivity which is a key element for the design of compact microwave detectors without the need of microwave filters.

We observed that the MTJ with 270 nm in diameter exhibited the largest ST-FMR signal. This is related to fact the breathing mode is related to the excitation of a skyrmion mode with size comparable to the MTJ cross section. In fact, the ST-FMR data in MTJs with smaller diameter that we have measured do not exhibit this breathing mode, at the same time, the ST-FMR signal measured in larger MTJs is smaller (see Supplementary Figure S7 for the characterization of magnetoresistance and ST-FMR data for MTJ devices having diameter of 180 and 420 nm).

Technologically, it is more relevant to design an STD at zero field in order to reduce both the energy consumption and improve the scalability and simplify the integration with semiconductor technology. The previous topological STD response is observed in many samples with the same nominal geometry, and, in some of them, also at zero field. The stabilization of zero field skyrmion is achieved via the application of a field protocol (see inset in **Figure 4**a) similar to Ref. [44]. The main panel of **Figure 4**a displays the corresponding magnetoresistance response. As the field decreases, the magnetoresistance at zero field decreases up to 140 Ω for field amplitudes lower than 10 Oe (see dark blue loop in **Figure 4**a). This resistance value is very close to that one at the first jump of the magnetoresistance ($H$=110 Oe) for the two loops with field amplitude larger than 10 Oe, thus pointing out the stabilization of a zero field skyrmionic state.

**Figure 4**b shows the zero field ST-FMR spectrum related to the skyrmionic state. We notice that it is qualitatively different from those under the presence of the field. In particular, here the rectified voltage due to the breathing mode exhibits both a positive (3 μV) and a negative peak (-1 μV) around 2 GHz. We ascribe this behavior to the change of the phase between the magnetoresistance signal and the injected ac current $I_{ac}$. The inset of **Figure 4**b shows the linear dependence of the maximum value of the positive peak of the rectified voltage vs the input rf power. The zero field ST-FMR spectra for different input rf powers are illustrated in Supplementary Figure S8.

### 2.3. MFM imaging of a single skyrmion in the patterned MTJ

To confirm that skyrmions are present within the MTJ, we performed OOP field MFM measurements (see Methods). **Figure 5**a and c show the sketch of the measured device at $H = 300$ and 1200 Oe, respectively, with the corresponding magnetization configuration in both the sample and the MFM tip. **Figure 5**b shows the hysteresis loop of an MTJ nominally the same as the one measured with MFM and the one measured in **Figure 3**, with a nominal diameter of 270 nm.

The contrast in **Figure 5**d is dominated by the north stray field from the upper SAF layer interacting with the north tip magnetization, resulting in an attractive tip-sample interaction and a negative (dark) frequency shift. The faint white ring arises from the return of the magnetic flux outside the MTJ. The stronger magnetic tip-sample interaction from the upper layer of the reference SAF, completely dominates the weaker magnetic contrast contribution from the SkyL located at the bottom of the MTJ stack. Indeed, unlike the hysteresis loop in **Figure 5**b, the raw MFM data recorded at field of 600 Oe displayed in **Figure 5**e did not show a noticeable change of the micromagnetic structure compared with that observed at 300 Oe displayed in **Figure 5**d. At $H = 1200$ Oe (**Figure 5**f), we still observe a similar contrast despite a clear change being shown in the hysteresis loop data. The only difference observed was the slightly more pronounced contrast of the white ring due to the lower layer within the SAF having been switched to north, aligning with the upper layer. Hence, imaging the relevant layer when located at the bottom of an MTJ is a challenging task for any method mapping the stray field above the surface, including MFM.

To overcome this issue, we employed a differential imaging technique where the contrast contribution from the SAF was subtracted from the raw data (see Methods). **Figure 5**g shows a homogenous south-oriented state of the SkyL at $H = 300$ Oe, as expected from the hysteresis loop in **Figure 5**b. At $H = 600$ Oe (**Figure 5**h), we observe a north-core skyrmion inside the MTJ, consistent with the hysteresis loop. We also observe north-oriented domains at the edges of the MTJ which are nucleated due to the IDMI boundary conditions[50]. When the field is further increased to $H = 1200$ Oe (**Figure 5**i), we see a homogeneous north-oriented state, confirming that the SkyL was saturated. Our processed MFM data thus proves the presence of a single skyrmion within the MTJs within the expected field range. It may seem convenient, and is often done, e.g. by Chen et al. [31], to apply a more binary contrast scale to create a more satisfying visualization of the micromagnetic state, as we have done in **Figure 5**j to k.

### 2.4. Micromagnetic simulations

To corroborate the experimental results and achieve a more detailed understanding of the modes excited in the MTJ, we perform micromagnetic simulations based on the numerical integration of the Landau-Lifshitz-Gilbert equation. This includes the STT applied to both the FL and RL of the MTJ and the layer dependence of the magnetic parameters to reflect the different material compositions of the stack (see Methods).

**Figure 6**a shows a comparison between the frequency modes measured experimentally (dashed lines), already shown in **Figure 3**c, and the results of micromagnetic simulations (solid symbols). For large OOP fields $H \geqslant 800$ Oe, both magnetizations of the FL and RL are uniform, with the one in the FL aligned along the positive z-axis parallel to the applied field, and the one of the RL pointing in the opposite direction (AP state). The micromagnetic simulations are in quantitative agreement with the experimental data and thus confirm the link of the HFM and LFM with the resonant mode of the FL and RL, respectively. In detail, the HFM with the blue-shift is excited into the FL, while the LFM exhibiting the red-shift is related to the RL dynamics.

At smaller field $200 < H < 600$ Oe, a single mode is observed. In order to understand its origin, we have studied the static characteristics of the skyrmionic state for a wide range of parameters. Experimentally (see **Figure 3**b) in this field region, both the uniform and the skyrmion state can be stabilized according to the field history. The theoretical details of the existence of this bi-stability region can be understood with theories published previously[51].

To describe the mode in region A, a tubular pure Néel skyrmion is set as the initial state in the SkyL (see **Figure 6**b), as suggested by the appearance of the intermediate states in the ascending branch of the hysteresis loop in **Figure 3**c. Whereas, FL, RL and SAF layers are considered in a uniform state. The calculation of the equilibrium configuration of the magnetization reveals that a Néel skyrmion is also nucleated in the CoFeB FL via the magnetostatic interaction from the skyrmionic state. This aspect is crucial to understand the reason why it can be detected electrically. This equilibrium state was used as the initial state for the micromagnetic frequency response analysis, where we use as input a microwave spin-polarized current. The micromagnetic simulations are again in agreement with the experimental results and show the excitation of the breathing mode also exhibiting a slight red-shift. Our calculations also show that the red shift is due to the interaction of the skyrmion domain wall with the MTJ boundary. To rule out the possibility that the mode in Region A could be associated with a different texture, i.e. a domain wall, we performed additional simulations with the same parameters. We observed that the excited resonance mode exhibits a qualitatively different trend on field, namely a slight blue-shift (see Supplementary Figure S9).

**Figure 6**c and d depict the magnetization amplitude as a function of the input frequency for the FL and RL in the uniform AP state (**c**) and with the FL in the skyrmionic state (**d**), respectively. The skyrmion breathing mode amplitude is slightly larger than the one of the uniform FL (black curve in d vs blue curve in **c**). However, the clear difference concerns the RL. The amplitude of this mode is much larger in the uniform precession, in agreement with the experimental data (a small amplitude RL mode in region A red curve in **d** which is observed in simulations can be also seen as weak mode in **Figure 3**c).

### 2.5. Comparison with previous works

The stabilization of skyrmions in an MTJ has been indirectly demonstrated a few years ago[23-24] and, more recently directly imaged at room temperature in thin-film MTJs[25, 30, 32] as well as in patterned MTJs[31, 52]. In those works, the skyrmionics state was manipulated by magnetic field or voltage controlled magnetocrystalline anisotropy (VCMA).

The first step to have miniaturized devices for the simultaneous magnetic skyrmion electrical excitation and imaging at room temperature relies on the design of an optimized material stack that includes (i) a multilayer hosting skyrmions, (ii) a large enough TMR, and (iii) a Resistance-Area (RA) product as low as possible. The latter is important to accommodate STT for the electrical excitation and detection of the skyrmion modes. In addition, to have the STT effect, the cross section of the device has to be small enough to have a negligible effect of the Oersted field generated by the current flowing in the device

We wish to highlight that it is a main challenge to reduce the RA of the MTJ and achieve a high TMR value at room temperature and that our MTJs have the smallest RA product up to date in MTJs hosting magnetic skyrmions, as summarized in the comparison shown in Table I .

| **Reference** | **MTJ size / TMR** | **RA** | **Topological texture (demonstration)** | **Manipulation** |
|---|---|---|---|---|
| Ref [23] | 250nm/ TMR <10% at 4K | 30 $\Omega\mu m^2$ | Single skyrmion (no image, TMR loop, micromagnetic simulations) | Nucleation field and microwave current |
| Ref [24] | 0.4 - 2 μm/ TMR <1% at 300 K | 20 k$\Omega\mu m^2$ | Single skyrmion (no images, TMR loop) | Nucleation and annihilation via VCMA pulses |
| Ref [25] | 3 - 10 μm, TMR 1.5% at 300 K | 80 k$\Omega\mu m^2$ for 5 μm MTJ | Multi Skyrmions (TMR loop, MFM images, micromagnetic simulations) | Field |
| Ref [30] | 2 μm, TMR 20% at 300 K | 291 k$\Omega\mu m^2$ | Multi Skyrmions (TMR loop, MFM images, micromagnetic simulations) | Field |
| Ref [32] | 500 nm, TMR 100% at 300 K (Extended free layer) | 2 k$\Omega\mu m^2$ | Multi Skyrmions (TMR loop) | Field |
| Ref [31] | 300 nm, TMR 20-70% at 300 K | 23 k$\Omega\mu m^2$ | Single Skyrmion (TMR loop, MFM images, micromagnetic simulations, zero field demonstration) | Nucleation and annihilation via VCMA pulses |
| Ref [52] | 500 nm, TMR 50% at 300 K | 700 $\Omega\mu m^2$ | Single Skyrmion (TMR loop, STXM images, zero field demonstration) | Nucleation and annihilation via VCMA pulses |
| This work | 270 nm, TMR 45% at 300 K | 12 $\Omega\mu m^2$ | Single Skyrmion (TMR loop, MFM images, STD detection, micromagnetic simulations, zero field demonstration) | Field and STT |

**Table 1.** A comparison of the MTJ characteristics and the methods for the detection and manipulation of magnetic skyrmions.

## 3. Summary and Conclusions

In summary, we have developed a topological STD with sub-300 nm MTJs combining a SkyL able to host skyrmions at room temperature and a TMR stack. ST-FMR measurements reveal the excitation of the breathing mode of a skyrmion that, for our devices, exhibits a resonance frequency near 4 GHz in a region of applied fields. We have also demonstrated the topological STD effect at zero field after applying a proper field protocol to stabilize zero-field skyrmions. Quantitative MFM measurements have been used to confirm the imprinting of the skyrmionic state from the SkyL into the MTJ FL. Micromagnetic simulations have supported quantitatively the experimental evidence.

Our work not only shows the proof of concept of the first topological STD and initiates the development of skyrmion-based microwave technology, but also opens up the accomplishment of promising perspectives. Technologically, the ST-FMR due to the skyrmion mode exhibits a narrower bandwidth compared to the uniform precession in the same device, thus suggesting a path toward the design of highly-selective microwave detectors. Our results also call for the integration of skyrmionic materials hosting smaller skyrmions (i.e. materials with larger DMI), with state-of-the-art TMR stacks for scaling the size of the topological diodes and enhancing the detection sensitivity by increasing the TMR ratio. The outcomes of this work could also positively impact the field of unconventional computing with skyrmions. At the basis of reservoir computing with skyrmions, there is the detection of skyrmion dynamics. For instance, in the approach suggested in Ref. [53], an MTJ could be implemented on top of a pinned skyrmion in order to detect its dynamics via the STD effect once excited by an in-plane electrical current. Similarly, our skyrmion based STD could be employed to detect the GHz dynamics of skyrmions in the task adaptive approach proposed in Ref. [54]. In neuromorphic computing, the ability of the skyrmion to detect a microwave input signal could impact the design of MTJs neural networks, where MTJs are used both as spin-torque oscillators and detectors to implement many-to-many connectivity through microwave signals[55]. The practical implementation of skyrmion quantum computing relies on experimental techniques able to identify the quantized helicity excitations of skyrmions via their dynamical response[56-57]. To this aim, our skyrmion based STD stands as a very promising solution and could move forward the field of skyrmion quantum computing.

This work is a breakthrough in the intriguing research direction aiming to move topological materials into device concepts and it is expected to have an impact on future engineering applications and developments in the field of spintronics and hybrid CMOS-spintronic technology.

**Methods**

*Sample preparation.* The skyrmionics MTJ devices studied here were patterned from a (thermal silicon oxide substrate)/ Ta (5)/CuN (20)/Ta (5)/[Pt (2.5)/Co (1)/Ta (0.5)]$_9$/Pt (2.5)/Co (1)/$Co_{40}Fe_{40}B_{20}$ (0.9)/MgO (0.85)/$Co_{20}Fe_{60}B_{20}$ (1.1)/Ta (0.5)/Co (0.3)/[Pt (1.5)/Co (0.4)]$_2$/Ru (0.85)/[Co (0.5)/Pt (1.5)]$_3$/Ru (5) multilayer, deposited by using a Singulus TIMARIS physical vapor deposition system (thickness in nm). The 0.5 nm thick Ta dusting layer was inserted between $Co_{20}Fe_{60}B_{20}$ and Co (0.3)/[Pt (1.5)/Co (0.4)]$_2$ layers to enhance the perpendicular magnetic anisotropy (PMA) of the top reference layers. Here a CuN layer was used as a buffer layer for the growth of the MTJ stack. This skyrmionic MTJ multilayer was post-annealed at 300 °C without a magnetic field for 1 hour to enhance the TMR ratio and PMA property. The films were subsequently patterned into nanopillars using optical and electron beam lithography combined with ion milling, incorporating an MgO barrier targeting a resistance-area product (R×A) of 12.3 $\Omega\mu m^2$ in the parallel magnetization configuration. The patterned electrodes, Ti (10 nm) and Au (100 nm), were fabricated using a lift-off process.

*Magnetoresistance and ST-FMR measurements.* All measurements reported in this paper are performed at room temperature. When the magnetoresistance was measured, a weak d.c. current $I_{dc}$ was applied to the device through a bias Tee using a source meter (2400, Keithley), the voltage of the device was recorded by a nanovolt meter (2182, Keithley). The out-of-plane TMR ratio, defined as $(R_{AP}-R_P)/R_P$, was 39.5%. When a microwave current $I_{ac}$ with a frequency $f_{ac}$ was applied to the device through a bias Tee using a signal generator (N5183B), the free layer magnetization starts to precess at the same frequency, resulting in a time-dependent resistance oscillation due to the TMR effect. As a result, a rectified voltage is generated across the MTJ. To improve the signal-to-noise ratio, the microwave input was modulated at a low frequency (97 Hz), and the resulting rectified voltage $V_{dc}$ was measured with a lock-in amplifier (SR830, Standard Research Systems).

*Magnetic force microscopy measurements and analysis.* The MFM measurements were performed under vacuum conditions (≈$10^{-6}$ mbar) in a home-built set up with an in-situ magnetic field supporting up to ±300 mT. Whilst in vacuum, the mechanical quality factor, Q, of the cantilever can exceed 200,000, which in comparison to MFM in ambient conditions, improves the measurement sensitivity forty-fold[46]. Furthermore, it allows the tip to be coated with only a thin magnetic layer, which minimizes the influence of its stray field on the samples micromagnetic state. The tips used were uncoated SS-ISC cantilevers from Team NanoTech GmbH of less than 5 nm in radius. The tips were sputtered at room temperature with a 2 nm Ta seed layer, a 4 nm Co layer (permitting the sensitivity to magnetic fields), and a 4 nm Ta cap to prevent oxidation. To oscillate the cantilever at a constant amplitude of 10 nm, and measure the frequency shift caused by the tip-sample interaction force derivative, a Zurich Instruments phase-locked loop (PLL) was used.

To perform the MFM measurements, we grew additional MTJs nominally the same as the one in **Figure 3** without the top gold contact. The MTJ was initially saturated in a south-oriented field, which meant at $H = 300$ Oe the SkyL was oriented south, the upper (lower) part of the SAF was north (south), and the tip was set to be north-orientated for the measurements.

MFM, which like any other method that measures a magnetic field, detects the sum of the stray fields coming from the individual layers within the MTJ, including the SAF. We must consider this when performing these measurements, and the consequences become apparent in the data displayed in **Figure 5**d, e, and f, showing the raw MFM data of the MTJ measured in different applied fields. Imaging the relevant layer when located at the bottom of an MTJ is a challenging task for any method mapping the stray field above the surface, including MFM. Without employing differential imaging techniques, it would not be possible to explore the micromagnetic structure of the layer of interest, located at the bottom of the MTJ.

To obtain the much weaker contrast from the SkyL, the dominating contrast contribution from the SAF must be subtracted from the raw data, i.e., differential imaging techniques must be employed. This requires that all data is acquired at exactly the same tip-sample distance. For this, all measurements were performed in constant height mode, i.e. not tracking the local topography. Note that this meant there was a reduced the tip-sample distance above the junction compared to the area surrounding the junction, and consequently, to a more attractive van der Waals interaction and a negative (dark) frequency shift over the junction. We used a capacitive tip-sample distance feedback method, discussed in our earlier work[47], to set a small average tip-sample distance in the area around the MTJ. For the successive image data acquisition, the z-feedback was stopped and the tip was retracted by a fixed distance to scan the MTJ at constant height.

*Micromagnetic simulations.* PETASPIN, an in-house CUDA-native full micromagnetic solver, was used to perform the micromagnetic simulations. This tool numerically integrates the Landau-Lifshitz-Gilbert (LLG) equation by applying the time solver scheme Adams-Bashforth[58]:

$$\frac{d\mathbf{m}}{d\tau} = -(\mathbf{m} \times \mathbf{h_{eff}}) + \alpha_G \left(\mathbf{m} \times \frac{d\mathbf{m}}{d\tau}\right) \quad (1)$$

where $\mathbf{m} = \mathbf{M} / M_S$ is the normalized magnetization, $\alpha_G$ is the Gilbert damping, and $\tau = \gamma_0 M_S t$ is the dimensionless time, which uses $\gamma_0$ the gyromagnetic ratio and $M_S$ the saturation magnetization. The normalized effective magnetic field, $\mathbf{h_{eff}}$, includes the exchange, interfacial DMI, magnetostatic, anisotropy and external fields.

The experimental MTJ is modelled as a circular stack with a diameter of 300 nm. The [Pt(2.5 nm)/Co(1 nm)/Ta(0.5 nm)]9 SkyL is simulated by nine repetitions of a 1 nm thick Co ferromagnet separated by a 3 nm thick Ta/Pt non-magnetic layer. The CoFeB FL (0.9 nm) and CoFeB (1.1 nm) in the RL are simulated by a 1 nm thick ferromagnetic layer each. These two CoFeB layers are separated by a 1 nm thick non-magnetic layer corresponding to the experimental MgO layer (0.85 nm). The CoFeB layer in the RL is in contact with the $[Co/Pt]_n/Ru/[Co/Pt]_m$ SAF layer. We modelled the latter as standard SAF composed of two 1 nm thick ferromagnetic layers separated by a 1 nm thick non-magnetic layer. The SAF stack is characterized by the interlayer exchange coupling (IEC) of the Ruderman–Kittel–Kasuya–Yosida (RKKY) type, with an effective field of $\boldsymbol{h}_{\mathrm{IEC},i} = \frac{J_{\mathrm{IEC}}}{\mu_0 M_{s,i}^2 t_{\mathrm{NM}}} \boldsymbol{m}_j$, [59-60] where $i$,$j$ are the indices of the top and bottom layers of the SAF, respectively, $J_{\mathrm{IEC}}$ is the IEC constant, and $t_{\mathrm{NM}}$ is the thickness of the Ru spacer. We apply an OOP external field $\boldsymbol{H}$ tilted of $\theta$ with the respect to the z-axis. We use a cuboidal discretization cell of $3 \times 3 \times 1$ nm$^3$.

For the excitation of the magnetization dynamics both in the FL and RL, we rely on the spin-transfer torque (STT) mechanism, which includes the effect of the back torque[61] on the RL magnetization due to an AC current $I_{ac} = I_{ac,0}\sin(2\pi f t + \varphi)$ of frequency $f$, phase $\varphi$ and amplitude $I_{ac,0}$. The STT term is added to Eq. (1) as:

$$\begin{cases} \boldsymbol{\tau}_{STT-FL} = \dfrac{gP\mu_B}{\gamma_0 e M_{s-FL}^2 V_{FL}} I_{ac}[\boldsymbol{m}_{\mathrm{FL}} \times (\boldsymbol{m}_{\mathrm{FL}} \times \boldsymbol{m}_{\mathrm{RL}})] \\ \boldsymbol{\tau}_{STT-RL} = -\dfrac{gP\mu_B}{\gamma_0 e M_{s-RL}^2 V_{RL}} I_{ac}[\boldsymbol{m}_{\mathrm{RL}} \times (\boldsymbol{m}_{\mathrm{RL}} \times \boldsymbol{m}_{\mathrm{FL}})] \end{cases} \quad (2),$$

where $\boldsymbol{m}_{FL}$ refers to the FL and $\boldsymbol{m}_{RL}$ to the RL. P is the spin-polarization equal to 0.66, $\mu_B$ is the Bohr magneton, $g$ is the Landè factor, $e$ the electron charge, and $V_{FL(RL)}$ is the volume of the FL (RL).

For the Co ferromagnets in the [Pt(2.5 nm)/Co(1 nm)/Ta(0.5 nm)]$_9$ SkyL, we considered $M_s$ = 900 kA/m[30], perpendicular anisotropy constant $K_u$ = 0.85 MJ/m$^3$, and interfacial DMI constant $D$ = 2.2 mJ/m$^2$.

For the CoFeB FL and RL, as well as for the SAF layers we fix $M_s$ = 1200 kA/m. The perpendicular anisotropy constants of the FL and RL are extracted by fitting the experimental frequency vs. field response in **Figure 3**c for the FL and RL uniform precessing (applied fields from 800 to 1000 Oe) by means of the well-known Kittel expression $f_{\mathrm{Kit}} = \frac{\gamma_0}{2\pi}\left(H_{\mathrm{ext}} + H_{K,\mathrm{eff}}\right)$, where $H_{K,\mathrm{eff}}$ is the effective magnetic anisotropy field. We obtain $K_{\mathrm{U,FL}}$ = 1.04 MJ/m$^3$ and $K_{\mathrm{U,RL}}$ = 0.99 MJ/m$^3$ for the FL and RL, respectively. The interfacial DMI of the FL is fixed to 1.1 mJ/m$^2$ (large enough to stabilize a Néel skyrmion), [62] while a zero DMI is considered for the RL and SAF layers. $K_{\mathrm{U,RL}}$ is also used for the perpendicular anisotropy constant of the SAF layers. In addition, we used an antiferromagnetic $J_{\mathrm{IEC}}$ = -0.5 mJ/m$^2$ which is sufficiently large to maintain the SAF in the antiferromagnetic state under the application of the OOP external field. All the simulations are performed at zero temperature.

## Acknowledgements

R.T., M.C., E.D., and G.F. thank the projects PRIN 2020LWPKH7 "The Italian factory of micromagnetic modelling and spintronics", PRIN20222N9A73 "SKYrmion-based magnetic tunnel junction to design a temperature SENSor—SkySens", funded by the Italian Ministry of Research, and the project number 101070287—SWAN-on-chip—HORIZON-CL4- 2021-DIGITAL EMERGING-01. R.T., M.C., E.D., and G.F. are with the Petaspin TEAM and thank the support of the PETASPIN association (www. petaspin.com). R.T. and M.C. acknowledge support from the Project PE0000021, "Network 4 Energy Sustainable Transition – NEST", funded by the European Union – NextGenerationEU, under the National Recovery and Resilience Plan (NRRP), Mission 4 Component 2 Investment 1.3 - Call for tender No. 1561 of 11.10.2022 of Ministero dell'Università e della Ricerca (MUR) (CUP C93C22005230007). Z.Z. would like to acknowledge the National Natural Science Foundation of China (No. 52371206, 12474127) and K. C. Wong Education Foundation (No. GJTD-2019-14). B. F. acknowledges support by the CAS Young Talent program and Gusu Leading Talents Program (ZXL2023172). We also thank the Truth Instruments Co.ltd for the help in MOKE Characterizations.

## Author contributions

B.F., Z.Z., G.F., X.Z., and W.J. designed the experiments. B.F., D.Z. and A.C. prepared the films, S.L. and Y.W. performed the devices fabrications, TEM and P-MOKE characterizations. Y.W., S.L. and B.F. did the electrical characterizations. H.J.H. and E.D. performed the MFM measurements and data analysis, and wrote the text related to the MFM results. R.T., M.C. and G. F. designed the micromagnetic solver. R. T., and M.C. carried out the micromagnetic simulations. B.F., Y.W., S.L., G.F., R.T., and Z.Z. analyzed the data. B.F., G.F., and R.T. wrote the manuscript with the help of Z.Z.. All the authors commented on the final version of the manuscript.

**References**

[1] G. Finocchio, F. Büttner, R. Tomasello, M. Carpentieri, M. Kläui, *J. Phys. D. Appl. Phys.* **2016**, 49.

[2] J. Q. Li, X. F. Zhou, J. Li, L. F. Che, J. Yao, G. McHale, M. K. Chaudhury, Z. K. Wang, *Sci. Adv.* **2017**, 3.

[3] A. Fert, N. Reyren, V. Cros, *Nat. Rev. Mater.* **2017**, 2.

[4] G. Finocchio, C. Panagopoulos, *Magnetic skyrmions and their applications*, Woodhead Publishing, **2021**.

[5] B. Göbel, I. Mertig, O. A. Tretiakov, *Phys. Rep.* **2021**, 895, 1.

[6] M. J. Gilbert, *Commun. Phys.* **2021**, 4.

[7] T. Dohi, R. M. Reeve, M. Kläui, *Annu. Rev. Condens. Matter Phys.* **2022**, 13, 73.

[8] J. Sampaio, V. Cros, S. Rohart, A. Thiaville, A. Fert, *Nat. Nanotechnol.* **2013**, 8, 839.

[9] R. Tomasello, E. Martinez, R. Zivieri, L. Torres, M. Carpentieri, G. Finocchio, *Sci. Rep.* **2014**, 4.

[10] W. J. Jiang, P. Upadhyaya, W. Zhang, G. Q. Yu, M. B. Jungfleisch, F. Y. Fradin, J. E. Pearson, Y. Tserkovnyak, K. L. Wang, O. Heinonen, S. G. E. te Velthuis, A. Hoffmann, *Science* **2015**, 349, 283.

[11] S. Woo, K. Litzius, B. Krüger, M. Y. Im, L. Caretta, K. Richter, M. Mann, A. Krone, R. M. Reeve, M. Weigand, P. Agrawal, I. Lemesh, M. A. Mawass, P. Fischer, M. Kläui, G. R. S. D. Beach, *Nat. Mater.* **2016**, 15, 501.

[12] W. J. Jiang, X. C. Zhang, G. Q. Yu, W. Zhang, X. Wang, M. B. Jungfleisch, J. E. Pearson, X. M. Cheng, O. Heinonen, K. L. Wang, Y. Zhou, A. Hoffmann, S. G. E. te Velthuis, *Nat. Phys.* **2017**, 13, 162.

[13] B. He, R. Tomasello, X. M. Luo, R. Zhang, Z. Y. Nie, M. Carpentieri, X. F. Han, G. Finocchio, G. Q. Yu, *Nano Lett.* **2023**, 23, 9482.

[14] J. Grollier, D. Querlioz, K. Y. Camsari, K. Everschor-Sitte, S. Fukami, M. D. Stiles, *Nat. Electron.* **2020**, 3, 360.

[15] K. M. Song, J. S. Jeong, B. Pan, X. C. Zhang, J. Xia, S. Cha, T. E. Park, K. Kim, S. Finizio, J. Raabe, J. Chang, Y. Zhou, W. S. Zhao, W. Kang, H. S. Ju, S. Woo, *Nat. Electron.* **2020**, 3, 148.

[16] S. Li, W. Kang, X. C. Zhang, T. X. Nie, Y. Zhou, K. L. Wang, W. S. Zhao, *Mater. Horiz.* **2021**, 8, 854.

[17] J. Zázvorka, F. Jakobs, D. Heinze, N. Keil, S. Kromin, S. Jaiswal, K. Litzius, G. Jakob, P. Virnau, D. Pinna, K. Everschor-Sitte, L. Rózsa, A. Donges, U. Nowak, M. Kläui, *Nat. Nanotechnol.* **2019**, 14, 658.

[18] K. Raab, M. A. Brems, G. Beneke, T. Dohi, J. Rothörl, F. Kammerbauer, J. H. Mentink, M. Kläui, *Nat. Commun.* **2022**, 13.

[19] T. Xu, Z. Chen, H. A. Zhou, Z. D. Wang, Y. Q. Dong, L. Aballe, M. Foerster, P. Gargiani, M. Valvidares, D. M. Bracher, T. Savchenko, A. Kleibert, R. Tomasello, G. Finocchio, S. G. Je, M. Y. Im, D. A. Muller, W. J. Jiang, *Phys. Rev. Mater.* **2021**, 5.

[20] A. O. Mandru, O. Yildirim, R. Tomasello, P. Heistracher, M. Penedo, A. Giordano, D. Suess, G. Finocchio, H. J. Hug, *Nat. Commun.* **2020**, 11.

[21] J. H. Liu, C. K. Song, L. Zhao, L. Cai, H. M. Feng, B. X. Zhao, M. Q. Zhao, Y. Zhou, L. Fang, W. J. Jiang, *Nano Lett.* **2023**, 23, 4931.

[22] S. Ikeda, K. Miura, H. Yamamoto, K. Mizunuma, H. D. Gan, M. Endo, S. Kanai, J.

Hayakawa, F. Matsukura, H. Ohno, *Nat. Mater.* **2010**, 9, 721.
[23] N. E. Penthorn, X. Hao, Z. Wang, Y. Huai, H. W. Jiang, *Phys. Rev. Lett.* **2019**, 122.
[24] S. Kasai, S. Sugimoto, Y. Nakatani, R. Ishikawa, Y. K. Takahashi, *Appl. Phys. Express* **2019**, 12.
[25] S. Li, A. Du, Y. D. Wang, X. R. Wang, X. Y. Zhang, H. Y. Cheng, W. L. Cai, S. Y. Lu, K. H. Cao, B. Pan, N. Lei, W. Kang, J. M. Liu, A. Fert, Z. P. Hou, W. S. Zhao, *Sci. Bull.* **2022**, 67, 691.
[26] C. Moreau-Luchaire, C. Moutafis, N. Reyren, J. Sampaio, C. A. F. Vaz, N. Van Horne, K. Bouzehouane, K. Garcia, C. Deranlot, P. Warnicke, P. Wohlhüter, J. M. George, M. Weigand, J. Raabe, V. Cros, A. Fert, *Nat. Nanotechnol.* **2016**, 11, 444.
[27] A. Soumyanarayanan, M. Raju, A. L. G. Oyarce, A. K. C. Tan, M. Y. Im, A. P. Petrovic, P. Ho, K. H. Khoo, M. Tran, C. K. Gan, F. Ernult, C. Panagopoulos, *Nat. Mater.* **2017**, 16, 898.
[28] W. J. Li, I. Bykova, S. L. Zhang, G. Q. Yu, R. Tomasello, M. Carpentieri, Y. Z. Liu, Y. Guang, J. Gräfe, M. Weigand, D. M. Burn, G. van der Laan, T. Hesjedal, Z. R. Yan, J. F. Feng, C. H. Wan, J. W. Wei, X. Wang, X. M. Zhang, H. J. Xu, C. Y. Guo, H. X. Wei, G. Finocchio, X. F. Han, G. Schütz, *Adv. Mater.* **2019**, 31.
[29] Z. D. Wang, M. H. Guo, H. A. Zhou, L. Zhao, T. Xu, R. Tomasello, H. Bai, Y. Q. Dong, S. G. Je, W. L. Chao, H. S. Han, S. Lee, K. S. Lee, Y. Y. Yao, W. Han, C. Song, H. Q. Wu, M. Carpentieri, G. Finocchio, M. Y. Im, S. Z. Lin, W. J. Jiang, *Nat. Electron.* **2020**, 3, 672.
[30] Y. Guang, L. K. Zhang, J. W. Zhang, Y. D. Wang, Y. L. Zhao, R. Tomasello, S. F. Zhang, B. He, J. H. Li, Y. Z. Liu, J. F. Feng, H. X. Wei, M. Carpentieri, Z. P. Hou, J. M. Liu, Y. Peng, Z. M. Zeng, G. Finocchio, X. X. Zhang, J. M. D. Coey, X. F. Han, G. Q. Yu, *Adv. Electron. Mater.* **2023**, 9.
[31] S. H. Chen, J. Lourembam, P. Ho, A. K. J. Toh, J. F. Huang, X. Y. Chen, H. K. Tan, S. L. K. Yap, R. J. J. Lim, H. R. Tan, T. S. Suraj, M. I. Sim, Y. T. Toh, I. Lim, N. C. B. Lim, J. Zhou, H. J. Chung, S. T. Lim, A. Soumyanarayanan, *Nature* **2024**, 627.
[32] M. Q. Zhao, A. T. Chen, P. Y. Huang, C. Liu, L. C. Shen, J. H. Liu, L. Zhao, B. Fang, W. C. Yue, D. X. Zheng, L. D. Wang, H. Bai, K. Shen, Y. Zhou, S. S. Wang, E. L. Liu, S. K. He, Y. L. Wang, X. X. Zhang, W. J. Jiang, *npj Quantum Mater.* **2024**, 9.
[33] J. L. Cai, B. Fang, L. K. Zhang, W. X. Lv, B. S. Zhang, T. J. Zhou, G. Finocchio, Z. M. Zeng, *Phys. Rev. Appl.* **2019**, 11.
[34] J. L. Cai, L. K. Zhang, B. Fang, W. X. Lv, B. S. Zhang, G. Finocchio, R. Xiong, S. H. Liang, Z. M. Zeng, *Appl. Phys. Lett.* **2019**, 114.
[35] N. Leroux, D. Markovic, E. Martin, T. Petrisor, D. Querlioz, A. Mizrahi, J. Grollier, *Phys. Rev. Appl.* **2021**, 15.
[36] G. Finocchio, M. Ricci, R. Tomasello, A. Giordano, M. Lanuzza, V. Puliafito, P. Burrascano, B. Azzerboni, M. Carpentieri, *Appl. Phys. Lett.* **2015**, 107.
[37] J. V. Kim, F. Garcia-Sanchez, J. Sampaio, C. Moreau-Luchaire, V. Cros, A. Fert, *Phys. Rev. B* **2014**, 90.
[38] M. Garst, J. Waizner, D. Grundler, *J. Phys. D. Appl. Phys.* **2017**, 50.
[39] Y. Onose, Y. Okamura, S. Seki, S. Ishiwata, Y. Tokura, *Phys. Rev. Lett.* **2012**, 109.
[40] T. Schwarze, J. Waizner, M. Garst, A. Bauer, I. Stasinopoulos, H. Berger, C. Pfleiderer, D. Grundler, *Nat. Mater.* **2015**, 14, 478.
[41] A. Aqeel, J. Sahliger, T. Taniguchi, S. Mändl, D. Mettus, H. Berger, A. Bauer, M. Garst,

C. Pfleiderer, C. H. Back, *Phys. Rev. Lett.* **2021**, 126.
[42] B. Satywali, V. P. Kravchuk, L. Q. Pan, M. Raju, S. He, F. S. Ma, A. P. Petrovic, M. Garst, C. Panagopoulos, *Nat. Commun.* **2021**, 12.
[43] T. Srivastava, Y. Sassi, F. Ajejas, A. Vecchiola, I. N. Yemeli, H. Hurdequint, K. Bouzehouane, N. Reyren, V. Cros, T. Devolder, J. V. Kim, G. de Loubens, *APL Mater.* **2023**, 11.
[44] N. K. Duong, M. Raju, A. P. Petrovic, R. Tomasello, G. Finocchio, C. Panagopoulos, *Appl. Phys. Lett.* **2019**, 114.
[45] S. J. Luo, N. Xu, Z. Guo, Y. Zhang, J. M. Hong, L. You, *IEEE Electron. Device Lett.* **2019**, 40, 635.
[46] Y. Feng, P. M. Vaghefi, S. Vranjkovic, M. Penedo, P. Kappenberger, J. Schwenk, X. Zhao, A. O. Mandru, H. J. Hug, *J. Magn. Magn. Mater.* **2022**, 551.
[47] X. Zhao, J. Schwenk, A. O. Mandru, M. Penedo, M. Bacani, M. A. Marioni, H. J. Hug, *New. J. Phys.* **2018**, 20.
[48] A. A. Tulapurkar, Y. Suzuki, A. Fukushima, H. Kubota, H. Maehara, K. Tsunekawa, D. D. Djayaprawira, N. Watanabe, S. Yuasa, *Nature* **2005**, 438, 339.
[49] G. Finocchio, R. Tomasello, B. Fang, A. Giordano, V. Puliafito, M. Carpentieri, Z. Zeng, *Appl. Phys. Lett.* **2021**, 118.
[50] S. Rohart, A. Thiaville, *Phys. Rev. B* **2013**, 88.
[51] R. Tomasello, K. Y. Guslienko, M. Ricci, A. Giordano, J. Barker, M. Carpentieri, O. Chubykalo-Fesenko, G. Finocchio, *Phys. Rev. B* **2018**, 97.
[52] J. U. Larrañaga, N. Sisodia, R. Guedas, V. Pham, I. Di Manici, A. Masseboeuf, K. Garello, F. Disdier, B. Fernandez, S. Wintz, M. Weigand, M. Belmeguenai, S. Pizzini, R. C. Sousa, L. D. Buda-Prejbeanu, G. Gaudin, O. Boulle, *Nano Lett.* **2024**, 24, 3557.
[53] D. Pinna, G. Bourianoff, K. Everschor-Sitte, *Phys. Rev. Appl.* **2020**, 14.
[54] O. Lee, T. Wei, K. D. Stenning, J. C. Gartside, D. Prestwood, S. Seki, A. Aqeel, K. Karube, N. Kanazawa, Y. Taguchi, C. Back, Y. Tokura, W. R. Branford, H. Kurebayashi, *Nat. Mater.* **2024**, 23, 79.
[55] A. Ross, N. Leroux, A. De Riz, D. Markovic, D. Sanz-Hernández, J. Trastoy, P. Bortolotti, D. Querlioz, L. Martins, L. Benetti, M. S. Claro, P. Anacleto, A. Schulman, T. Taris, J. B. Begueret, S. Saïghi, A. S. Jenkins, R. Ferreira, A. F. Vincent, F. A. Mizrahi, J. Grollier, *Nat. Nanotechnol.* **2023**, 18, 1273.
[56] C. Psaroudaki, C. Panagopoulos, *Phys. Rev. Lett.* **2021**, 127.
[57] C. Psaroudaki, E. Peraticos, C. Panagopoulos, *Appl. Phys. Lett.* **2023**, 123.
[58] A. Giordano, G. Finocchio, L. Torres, M. Carpentieri, B. Azzerboni, *J. Appl. Phys.* **2012**, 111.
[59] O. Alejos, V. Raposo, L. Sanchez-Tejerina, R. Tomasello, G. Finocchio, E. Martinez, *J. Appl. Phys.* **2018**, 123.
[60] E. Darwin, R. Tomasello, P. M. Shepley, N. Satchell, M. Carpentieri, G. Finocchio, B. J. Hickey, *Sci. Rep.* **2024**, 14.
[61] G. Siracusano, G. Finocchio, I. N. Krivorotov, L. Torres, G. Consolo, B. Azzerboni, *J. Appl. Phys.* **2009**, 105.
[62] A. Cao, X. Y. Zhang, B. Koopmans, S. Z. Peng, Y. Zhang, Z. L. Wang, S. H. Yan, H. X. Yang, W. S. Zhao, *Nanoscale* **2018**, 10, 12062.

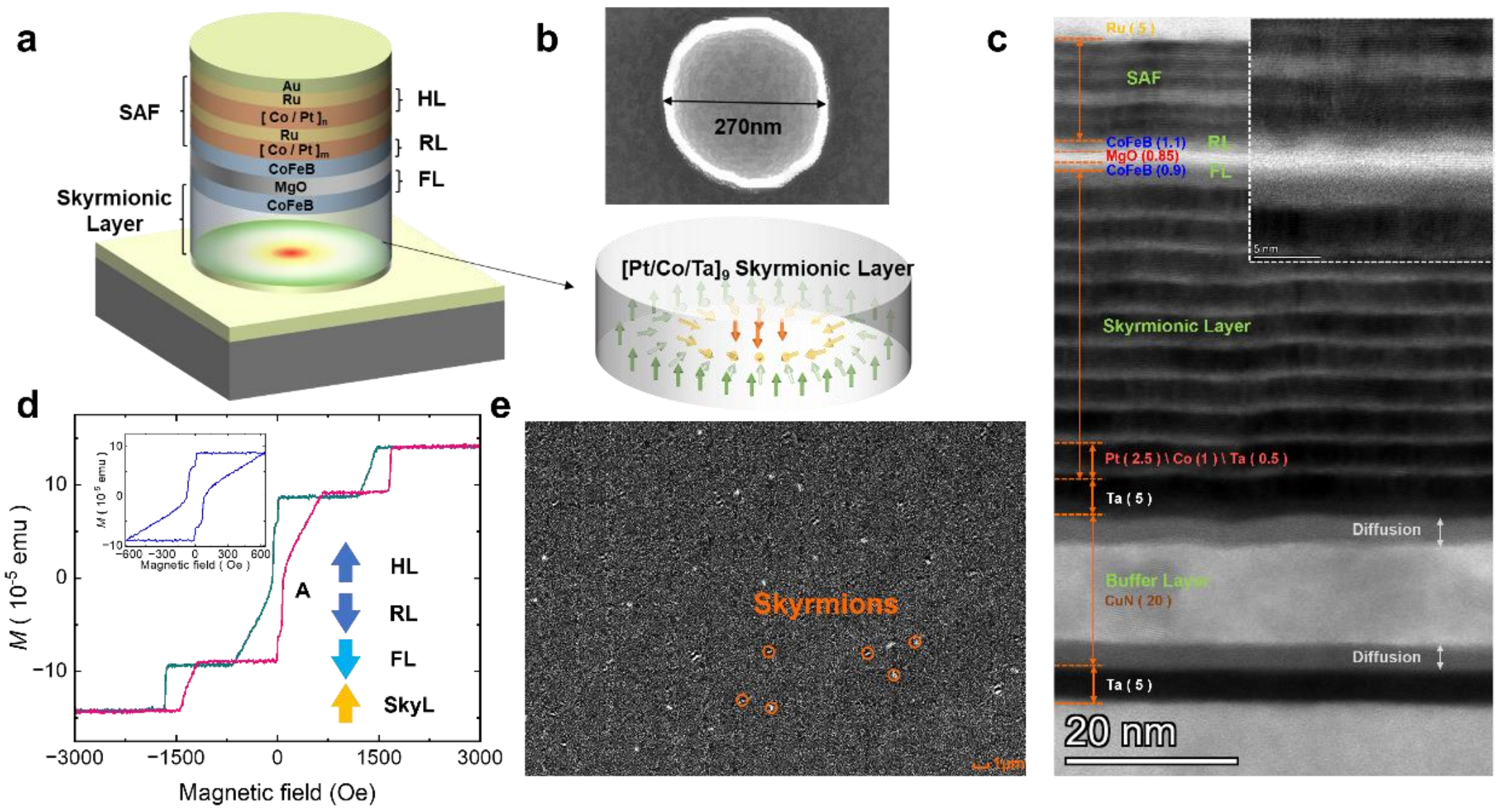

**Figure 1. Device stack and experimental characterization.** a) Schematic configuration of the device with the description of the films stack. The skyrmionic layer (SkyL), composed of $[Pt/Co/Ta]_9$, allows for the stabilization of magnetic skyrmions at room temperature. The CoFeB of the MTJ acting as the free layer (FL) is on top of the SkyL, while the reference layer (RL) is exchange biased by a SAF layer, HL is the hard layer. Both the FL and RL have a perpendicular easy axis. b) An SEM image of a patterned MTJ with a diameter of 270 nm. c) High resolution TEM (HRTEM) images of the whole stack. The HRTEM shows smooth interfaces with an indication of the thickness of each layer (in nm). The inset is a magnification of the MgO barrier layer and points out its high crystalline quality. d) OOP magnetic field hysteresis major loop measured by vibrating sample magnetometry (VSM). The inset shows the corresponding minor loop. e) Polar MOKE magnetic imaging of the $Ta/MgO/CoFeB/Ta/[Pt/Co/Ta]_9$ stacks for an OOP field of 200 Oe showing isolated skyrmions. The OOP magnetization component is color coded: white is in the +z orientation. Scale bar, 1 μm.

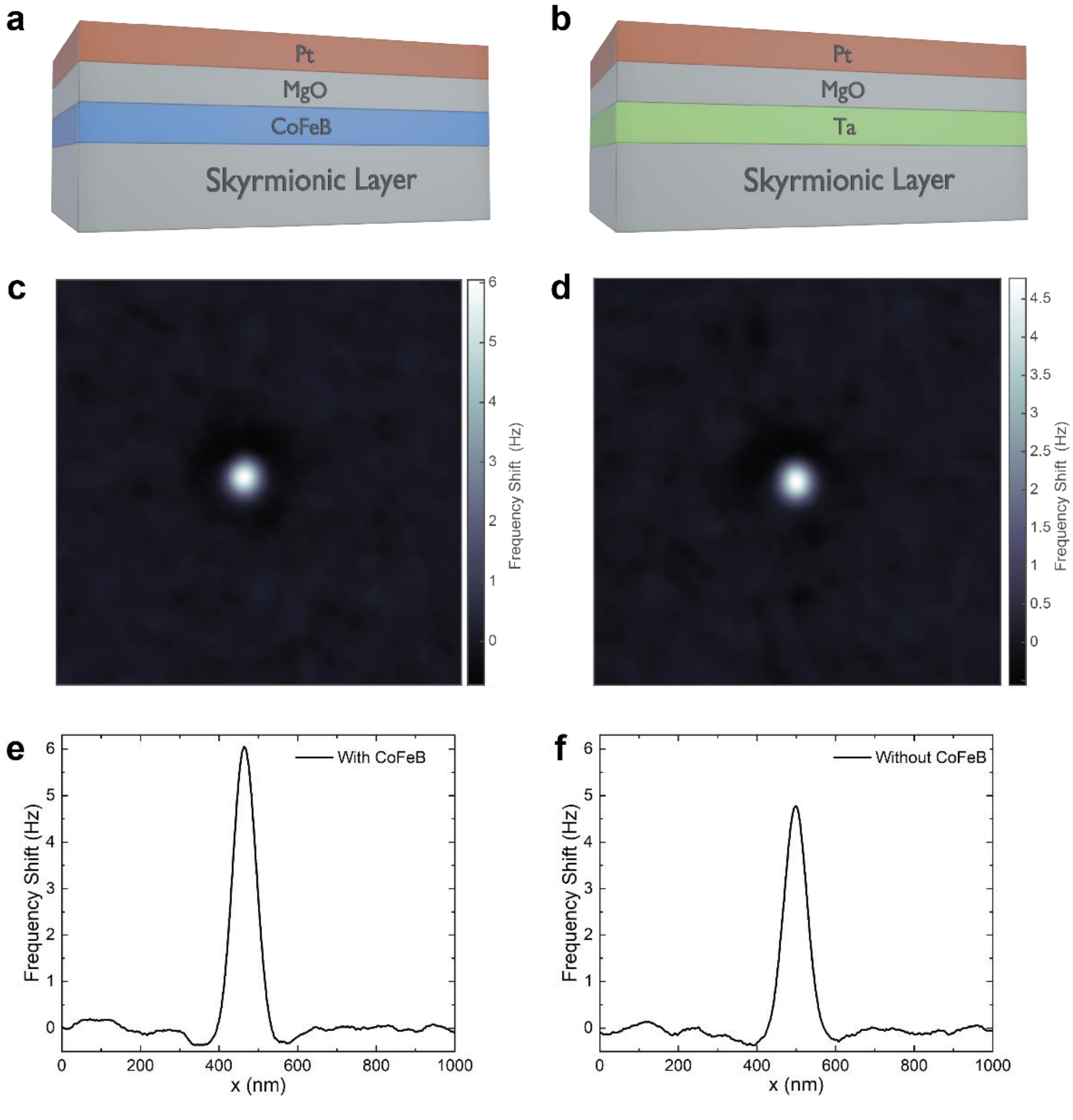

**Figure 2. MFM measurement of a skyrmion in the CoFeB layer.** a) Schematic of a multilayer, consisting of lower part of the MTJ stack up to the MgO, but without the reference layer, with a cap of Pt. b) The same as (a), but with the CoFeB layer replaced by a Ta layer of the same thickness. c) MFM frequency shift data of a skyrmion in the multilayer in (a), acquired in a field of $H$ = 1240 Oe. d) MFM frequency shift data of a skyrmion in the multilayer in (b), acquired in a field of $H$ = 1180 Oe. e) Line profile of the skyrmion in (c). f) Line profile of the skyrmion in (d). Note that the higher MFM contrast observed for sample (a), proves that the skyrmion also exists in the top CoFeB layer.

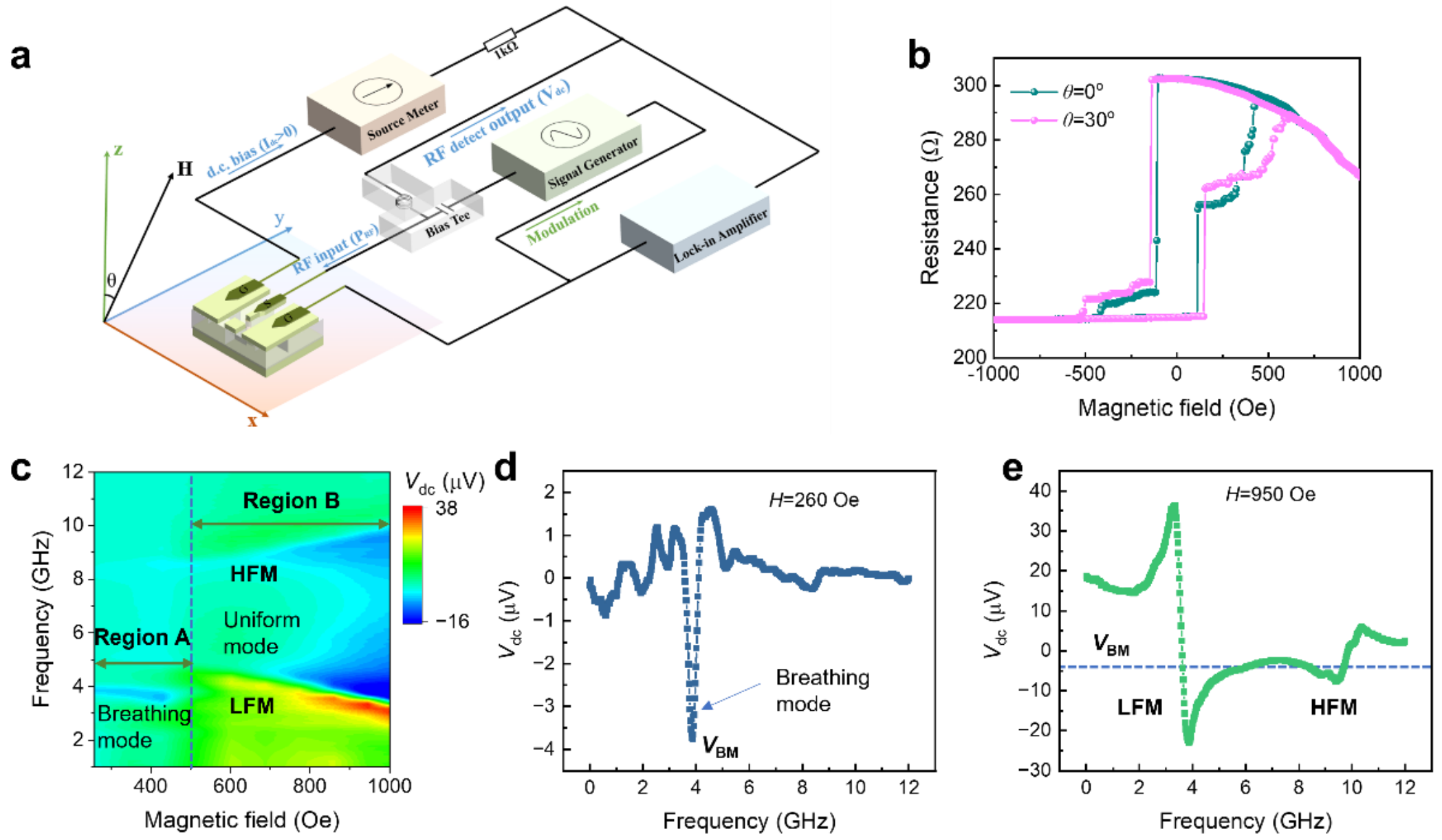

**Figure 3. Magneto-electrical transport and ST-FMR characterization of the skyrmionic MTJ.** a) Schematics of the circuit for the ST-FMR characterization, where a d.c. bias current is injected into the MTJ device by a DC terminal of the bias Tee. The differential voltage is recorded by the lock-in amplifier. For the ST-FMR measurement, an RF current is injected by the signal generator, the rectified voltage is also recorded by a low frequency (97 Hz) modulation method using the lock-in amplifier. b) Magnetoresistance data measured as a function of magnetic field amplitude at $\theta = 0°$ and $\theta = 30°$, respectively, where $\theta$ is the angle between the normal and the magnetic field defined in panel (a). The sharp change in the magnetoresistance curves indicates a switching process in the FL of the MTJ, while the gradual change can be linked to the skyrmionic region as observed in **Figure 1**d. c) ST-FMR spectra as a function of the amplitude of the external field for $\theta = 30°$ and $P_{rf} = 5\mu W$. The color code is linked to the value of the rectification voltage $V_{dc}$. Two regions are observed, region A is characterized by the excitation of the skyrmion breathing mode, while in region B, the excitation of the uniform modes for the FL and RL can be observed. d) and e) Examples of ST-FMR measurements as a function of the microwave frequency ($\theta = 30°$ and $P_{rf} = 5\mu W$) for region A at $H = 260$ Oe and region B at $H = 950$ Oe, respectively.

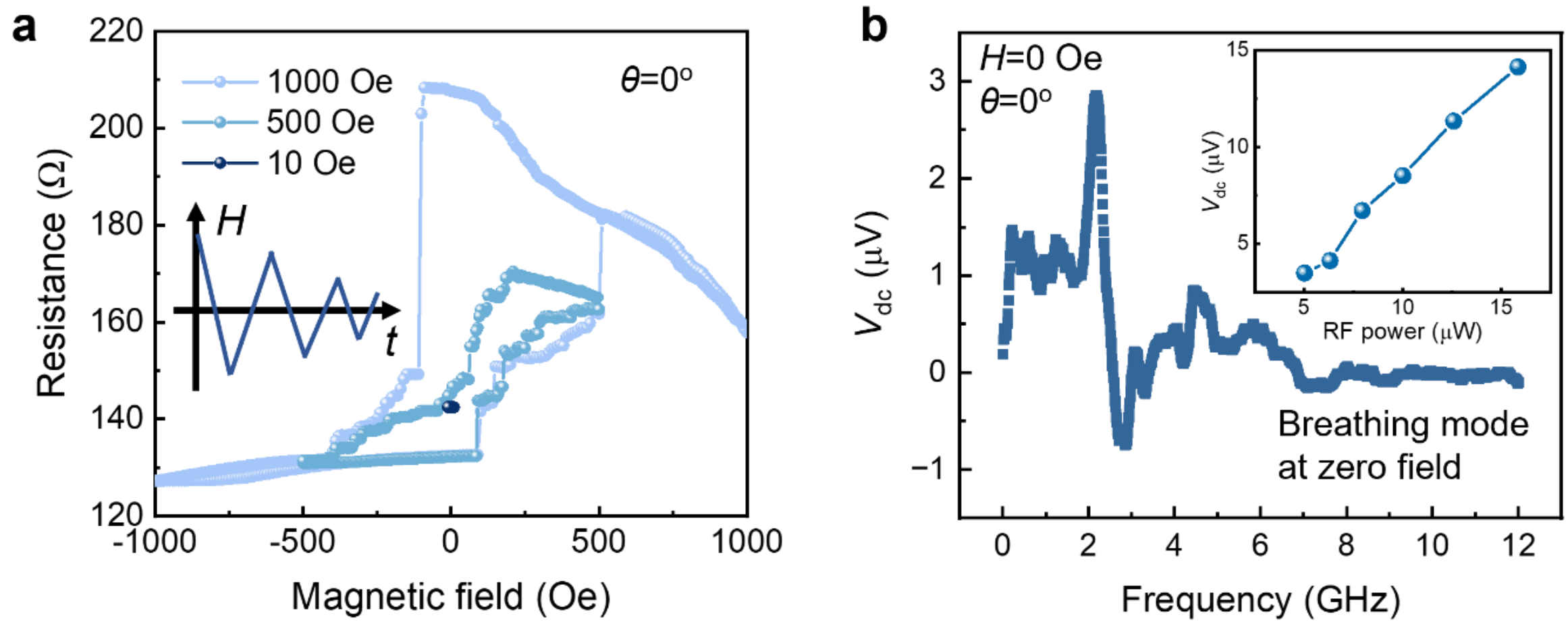

**Figure 4. Zero field skyrmionic state.** a) Magnetoresistance data measured as a function of the field protocol (a time-varying magnetic field) necessary to achieve a zero field skyrmionic state at $\theta = 0°$. The inset shows the qualitative time-dependence of the magnetic field. b) Zero field ST-FMR measurements as a function of the microwave frequency ($\theta = 0°$ and $P_{rf} = 5$ μW). The inset shows the maximum value of the positive peak of the rectified voltage as a function of the input rf power.

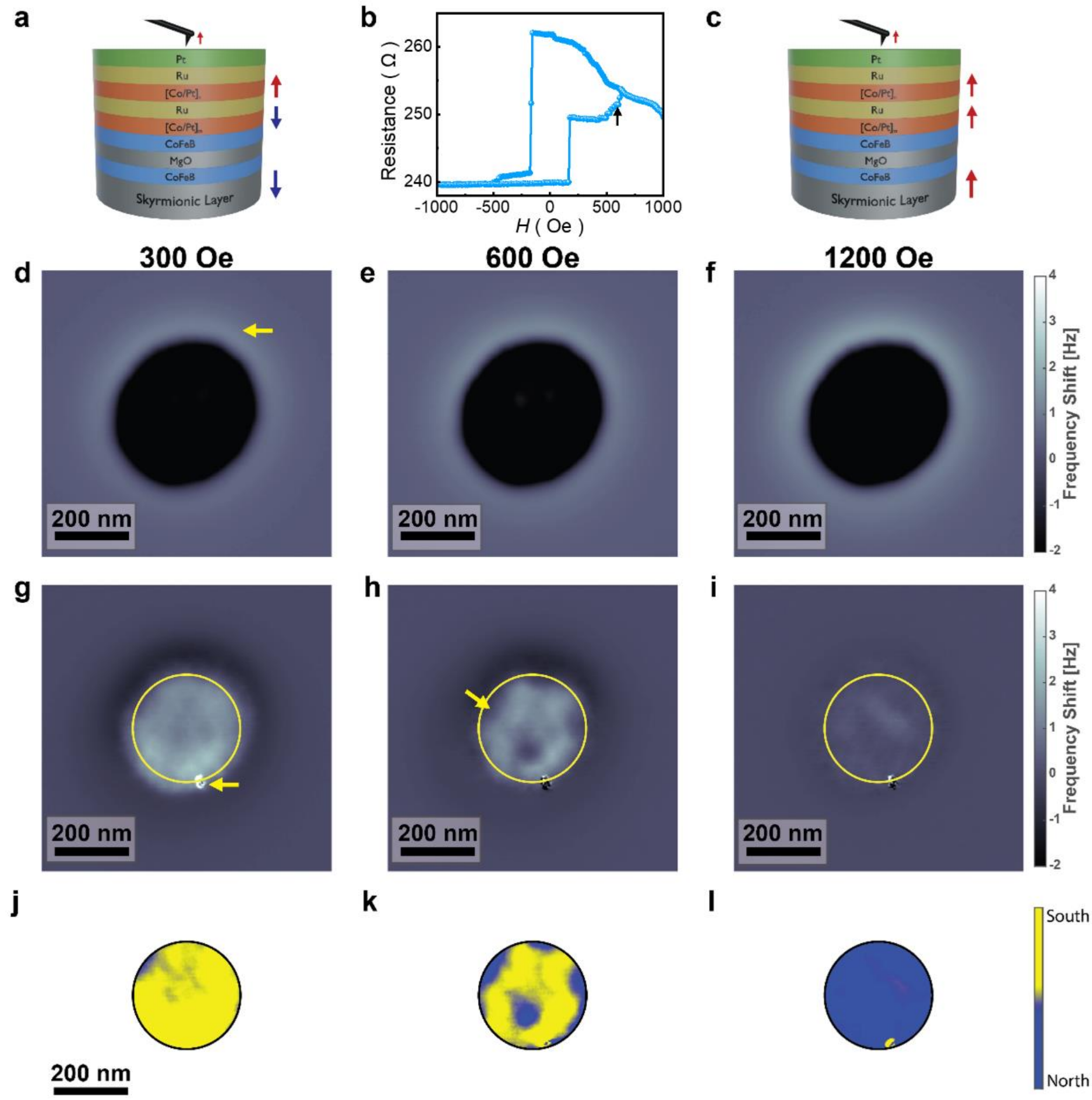

**Figure 5. MFM measurements of MTJs hosting a single skyrmion.** a) Schematic of the MTJ device measured by MFM, depicting the cantilever-tip above the sample surface and showing the magnetization orientations of the tip and layers at $H = 300$ Oe. The red up-arrows represent a north orientation, and the blue down-arrows represent a south orientation. b) Magnetoresistance data measured as a function of magnetic field for an MTJ nominally identical to the one measured by MFM. The black arrow points out 600 Oe. c) Schematic of the MTJ device measured by MFM, depicting the cantilever-tip above the surface showing the orientations of the tip and layers at $H = 1200$ Oe. d) to f) Raw MFM data of the MTJ with an applied OOP magnetic field of $H = 300$, 600 and 1200 Oe, respectively. The yellow arrow points to the white ring around the MTJ device, arising from the return of the magnetic flux outside the MTJ. The interaction of the stray field of the upper SAF reference layer with a north magnetization with the north-oriented tip results in an attractive tip-sample interaction force

derivative (negative frequency shift). This dominant contrast masks contributions from the lower SkyL. g) to i) Same images as in (d)-(f), but with the contribution from the SAF subtracted, allowing the weaker contrast contributions from the SkyL to be observed. Note that the SkyL has an initial south-oriented magnetization, leading to a positive frequency shift contrast when imaged with a north-oriented tip. The yellow circle encompasses the MTJ, and its slightly wider appearance in the MFM data is ascribed to the contrast generated from the MTJ's slanted side walls. In (g), the arrow points to a particle on the MTJ, which should not be mistaken for a magnetic texture, and in (h), the arrow points to an example of the canted edge states. j) to l) Same data as (g)-(i), however, displayed with a more simplified, almost binary color scale.

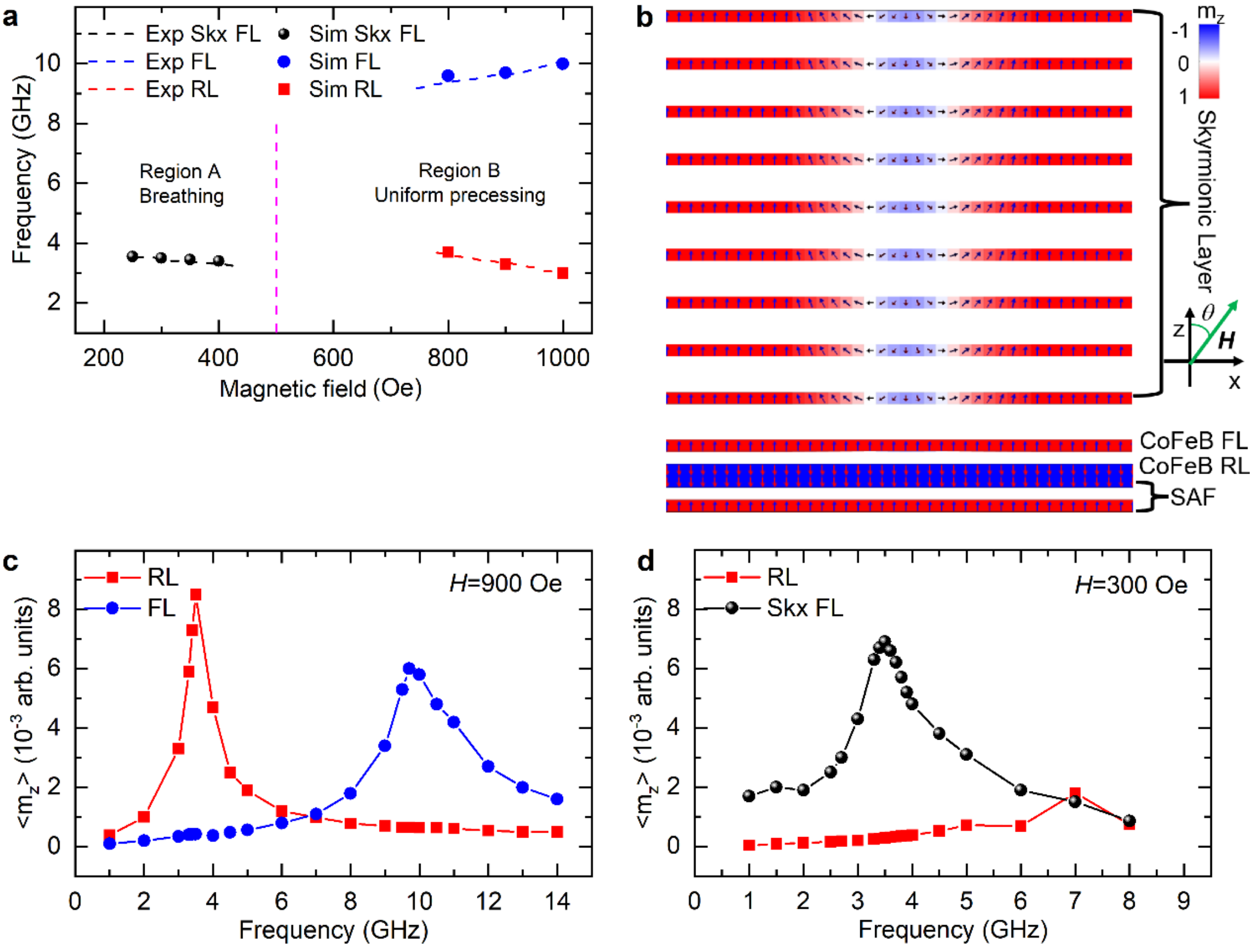

**Figure 6. Micromagnetic simulations results.** a) Frequency of the excited modes as a function of the applied field ***H*** for $\theta$ = 30°. Solid symbols represent the results of the micromagnetic simulations, dashed lines correspond to the experimental results extracted from **Figure 3**c. b) Cross-section of the simulated MTJ with the SkyL, where the initial state is a tubular Néel skyrmion (the color bar indicates the OOP component of the magnetization, the arrows indicate the direction of the magnetization in the x-z plane). The magenta dashed line divides region A where the skyrmion breathing mode occurs, from region B, where the uniform precession of both FL and RL is excited. c) Example of the frequency response for the FL and RL in the uniform AP state for *H*=900 Oe. d) Example of the frequency response for the FL and RL in the skyrmionic state for *H*=300 Oe.